\documentstyle{article}

\title{
\begin{flushright}
{\bf\normalsize   COLO-HEP-285}\\
\end{flushright}
\bf An Effective Model for Crumpling in
Two Dimensions?}
\author{ {\it C.F. Baillie} \\
         Physics Dept. \\
         University of Colorado\\
         Boulder, CO 80309\\
         USA\\
         \\
         and \\
         \\
         {\it D.A. Johnston}\\
         Dept. of Mathematics\\
         Heriot-Watt University\\
         Riccarton\\
         Edinburgh, EH14 4AS\\
         Scotland}
  
\begin{document}
  \maketitle
%-----------------------------------------------------------------------
                      {\Large
                      \begin{abstract}
%-----------------------------------------------------------------------
%
We investigate the crumpling transition for a dynamically triangulated
random surface embedded in two dimensions using an effective model
in which the disordering effect of the $X$ variables on the correlations of the
normals
is replaced by a long-range ``antiferromagnetic'' term.
We compare the results from a Monte Carlo simulation
with those obtained for the standard
action which retains the $X$'s and discuss the nature of the phase transition.
%
%-----------------------------------------------------------------------
                        \end{abstract} }
%-----------------------------------------------------------------------
%
  \thispagestyle{empty}
%
%***********************************************************************
%
  \newpage
%
%-----------------------------------------------------------------------
                  \pagenumbering{arabic}
%-----------------------------------------------------------------------
It is now clear that the pathologies of the continuum Polyakov action
for the bosonic string \cite{1}
in physical dimensions ($d>1$)
\begin{equation}
Z = \int D g D X \exp \left( -{1 \over 2} \int d^2 x \sqrt{g} g^{ab}
\partial_a X^{\mu} \partial_b X_{\mu} \right)
\label{e1}
\end{equation}
are also manifest in the discretized Euclidean form of the action used
in simulations
\begin{equation}
Z =  \sum_T \int \prod_i DX_i \exp \left( - { 1 \over 2}
    \sum_{<ij>} \left( X^\mu_i -X^\mu_j \right)^2 \right).
\label{e2}
\end{equation}
In this simple gaussian action
the continuum integration over the metric is replaced by a summation
over triangulations of the discretized worldsheet, the variables $X_i$ which
specify
the embedding of the surface typically live at
the nodes of the triangulation and the sum is over the edges $<ij>$.
Both microcanonical simulations, in which the number of nodes was fixed,
and grand canonical simulations, in which it varied, were carried out
and it was discovered that
the worldsheets
generated by the action were very crumpled and failed to give a
non-trivial continuum limit \cite{2}.

A possible remedy, suggested by Polyakov in \cite{3}, was the inclusion of
a stiffness term in the action which could be written using
either of the two equivalent forms of the extrinsic
curvature squared
\begin{eqnarray}
K^2 &=&  \int d^2x \sqrt{h}
\left( \Delta_h X^\mu \right)^2 \nonumber \\
    &=&  \int d^2 x \sqrt{h} h^{ab}
\nabla_a \hat n^\mu_i \nabla_b
\hat n_\mu^i.
\label{e3}
\end{eqnarray}
Subsequent simulations revealed that discretizations of the first version
were also afflicted
by pathologies \cite{4}
but that a discretization of the second form
\begin{equation}
S_e = \sum_{\Delta_i, \Delta_j} ( 1 - \hat n_{\Delta_i}
 . \hat n_{ \Delta_j} ),
\label{e5}
\end{equation}
where the $\Delta$'s are adjacent triangles and the $\hat n$'s normals,
offered the possibility of defining a continuum limit
at a second order phase transition where the string tension scaled
\cite{5}. Some recent simulations have cast doubt on the second order nature
of the transition, but it appears that the requisite scaling of the string
tension may still be taking place \cite{6}. Other alternative
actions are also being investigated \cite{6a}.

The extrinsic curvature squared term in eq.\ref{e5} may also be written as
\begin{equation}
S_e = \sum_{\Delta_i, \Delta_j} ( 1 - \cos \theta_{ij}),
\label{e6a}
\end{equation}
where the $\theta_{ij}$ is the angle between the adjacent normals.
In the degenerate case of a surface embedded in just two dimensions
it is still possible to write this term using
Ising-like variables $\sigma = \pm 1$
\begin{equation}
S_e = \sum_{<ij>} ( 1 - \sigma_i \sigma_j)
\label{e7}
\end{equation}
as the cosine will always take the value $\pm 1$.
(Note that these Ising variables live on the dual of the triangulation.)
This prompted Renken and Kogut
to suggest that if one ignored the effect of the $X$
variables in the gaussian term one might expect an Ising like crumpling
transition in
this case as $S_e$ in eq.\ref{e7} is essentially
the $2d$ Ising model action. This transition would thus be second order
and might serve as a model for the transition in
more realistic dimensions. However, their numerical simulations
of a gaussian plus extrinsic curvature squared action on a fixed triangulation
using finite size scaling techniques
revealed a first order transition \cite{7}. It does appear that this model
on a {\it dynamical} triangulation has a second order crumpling transition
which is more in line with expectation,
though it is only possible to work with much smaller meshes in this case
\cite{8}.

As we have ignored the gaussian term it is natural to enquire what effect its
inclusion might have on the analysis above. It was pointed out in \cite{9}
that if one used a proper time representation for the $X$ propagator
of a gaussian surface action
\begin{equation}
<X^{\mu} ( \xi_i ) X^{\nu} ( \xi_j ) > = \delta^{\mu \nu} \int^{\Lambda^2}_{1
\over L^2} { d \alpha \over 4 \pi \alpha} \exp ( - \alpha (\xi_i - \xi_j)^2 ),
\label{e11}
\end{equation}
where $L$ is an infrared cutoff and $\Lambda$ an ultraviolet cutoff,
then the normal-normal correlations were of the form
\begin{equation}
<n ( \xi_i ) \cdot n ( \xi_j ) >  \ \simeq\  - { A \over ( \xi_i - \xi_j)^4}.
\label{e12}
\end{equation}
The constant $A$ depends on the dimension in which the surface is embedded
\footnote{The considerations of \cite{9} were for {\it rigid} or crystalline
surfaces
but they are essentially unchanged for dynamical surfaces.}. From eq.\ref{e12}
it is clear that the gaussian term induces an antiferromagnetic
interaction between the normals.
This suggests a possible interpretation of
the crumpling transition as arising from the competition between the
gaussian ``antiferromagnetic'' term and the extrinsic curvature
``ferromagnetic''
term, which tends to order the normals.

In this paper we test this interpretation for a surface embedded in two
dimensions
by comparing the behavior of the ``effective'' action
where the gaussian term has been replaced by a term mimicing its effect on the
normal spins,
\begin{equation}
S = A \sum_{ij} { 1 \over r_{ij}^4} \sigma_i \sigma_j
+ \lambda \sum_{<ij>} ( 1 - \sigma_i \sigma_j),
\label{e13}
\end{equation}
with the standard action we simulated in \cite{8}
\begin{equation}
S =  {1 \over 2} \sum_{<ij>} (X_i^{\mu} - X_j^{\mu})^2
+ \lambda \sum_{<ij>} ( 1 - \sigma_i \sigma_j)
\label{e14}
\end{equation}
where $\lambda$ is the ferromagnetic coupling proportional to the inverse
temperature.
For convenience we set the constant $A=1$ in the simulations.
Note that the sum in the first term of eq.\ref{e13} is over {\it all} the spins
so we have a long-range interaction.
Moreover $r_{ij}$ is the {\it intrinsic} distance between the spins
$\sigma_i$ and $\sigma_j$, i.e. the shortest path (number of links)
between points $i$ and
$j$ in the triangulation. In fact since there are no $X$ variables in
eq.\ref{e13}, the triangulation is no longer embedded in any dimension
and there are no extrinsic distances. This means that we can no longer
measure the radius of gyration to see how the size of the surface is changing,
though we can measure the contribution of the extrinsic curvature squared term
eq.\ref{e7}
which we would expect to be small in a smooth phase and large in a crumpled
phase.

The action in eq.\ref{e13} is rather similar to that
suggested in \cite{10} to explain the phenomenon of spin canting and reentrance
\footnote{Reentrance is the behavior displayed by metallic spin glasses such as
AuFe
which have a paramagnetic to ferromagnetic transition at some $T_c$ and a
ferromagnetic to spin-glass transition at some $T_f<T_c$. Below $T_f$ spins
tend to rotate away from an external field, which is termed canting.}
\begin{equation}
S = - J_0 \sum_{<ij>} \vec S_i \cdot \vec S_j + 2 \epsilon \sum_{ij} {1 \over
r_{ij}^{d+\sigma}}\vec S_i \cdot \vec S_j
\label{e16}
\end{equation}
where the $\vec S$'s are classical two-component spins and the sum in the
second term is restricted to $r_{ij}<R$,
where $R$ is some cutoff. The authors in \cite{10} were interested in $d=3$ and
$\sigma<2$ and argued that
the competition between the short range $J_0$ term and the long-range
$\epsilon$ term accounted for canting.
Actions of the form
\begin{equation}
S = - J_0 \sum_{ij} {1 \over r_{ij}^{\alpha}} \sigma_i \sigma_j
\label{e17}
\end{equation}
have also been considered \cite{11}, and display phase transitions for
$d<\alpha<2d$.
In our case the $1/r^4$ interaction induced
between all the normals is apparently
on the borderline $\alpha = 2d$
if we assume $d=2$, but direct measurement
of $d$ on the highly irregular meshes generated by
2d quantum gravity that we use gives $d>2$. However our interaction is {\it
anti}ferromagnetic,
so it appears that the analogy with
the canting action is closer as we have two competing terms, one
``ferromagnetic'' and one ``antiferromagnetic'',  and observe a transition as
we vary their
relative strengths.

The most difficult part in performing a Monte Carlo simulation of the action
in eq.\ref{e13} is calculating the intrinsic distance $r_{ij}$ (represented
as a matrix)
between all points $i,j$ on the random triangulation.
There is a straight-forward algorithm which does this but it is $O(N^2)$,
where $N$ is the number of points in the triangulation.
This makes it prohibitively expensive to simulate {\it dynamical}
random triangulations, since after every change to the triangulation
one would have to recalculate $r_{ij}$
(even though the changes to the triangulation would consist only of
the standard ``flip'' move which is local this would still produce
non-local changes in the intrinsic distance matrix $r_{ij}$).
Therefore all our simulations use {\it fixed} random triangulations
so that we have only to calculate $r_{ij}$ once at the beginning.
As our model is essentially an Ising model coupled to two-dimensional
quantum gravity, albeit with an extra long-range term, we have
chosen to use random triangulations coming from pure two-dimensional
quantum gravity simulations as was done in \cite{12}.

We have performed simulations on three sizes of random triangulation
of spherical topology with total number of points $N = 100,500$ and $1000$.
For each $N$ we ran at roughly $20$ values of $\lambda$ between $1$ and
$3$. After thermalizing we do $100000$ Monte Carlo updates using the standard
Metropolis algorithm, measuring the energy, magnetization and spin-spin
correlation function after every update
which allows us to calculate autocorrelation times and correlation length.
We find, as expected for the local Metropolis algorithm, that the
autocorrelation time $\tau$ scales as the square of the correlation length
$\xi$; in fact for the energy, by fitting $\tau \sim \xi^z$, we extract
a dynamical critical exponent $z=2.1(2)$.
{}From the fluctuations in the energy and magnetization we obtain the
specific heat and susceptibility in the usual manner.

In Fig. 1 we show the total energy $E$ from the largest system
simulated $N=1000$ (graphs from $N=100$ and $500$ look almost identical)
and its two parts --
$E_{lr}$ is the antiferromagnetic long-range part (first term in eq.\ref{e13}
with A=1) and
$E_{nn}$ is the ferromagnetic nearest-neighbor part (second term in
eq.\ref{e13}
including $\lambda$).
In Fig. 2 we show the total magnetization $M$, from all three simulations.
At small $\lambda$, which corresponds to high temperature, we see from Fig. 1
that most
of the energy is in the ferromagnetic piece $E_{nn}$ since the spins are
disordered, which we would expect for a ``crumpled'' phase -- $M$ is small in
Fig. 2.
Conversely, at large $\lambda$ (low temperature) the spins line up so
$M \rightarrow 1$ and most of the energy resides in the antiferromagnetic
$E_{lr}$. In this smooth phase the contribution from the
extrinsic curvature squared $E_{nn}$ is small.
{}From these we deduce there is some sort of crumpling transition at
around $\lambda = 2.3 - 2.4$.
Also from Fig. 2 we see that as the system size increases the ``jump''
in $M$ becomes more abrupt, perhaps
signaling the appearance of a first order phase transition.
However there is no corresponding jump in the total energy so this seems
unlikely.

Stronger evidence of the phase transition not being of first
order comes from the specific heat, shown in Fig. 3.
For a first order phase transition, standard finite-size scaling theory
predicts that for a two-dimensional system of size $N=L \times L$, the
specific heat peak
\begin{equation}
C_{max} = A L^2 +B,
\label{e19}
\end{equation}
whereas for a second order transition
\begin{equation}
C_{max} = A' L^{\alpha/\nu} +B'.
\label{e20}
\end{equation}
We do not know {\it a priori} that our system is two-dimensional so we
must write $L=N^{1 \over d}$, where $d$ is the fractal dimension of the
random triangulation.
In fact, numerical simulations for pure 2d quantum gravity \cite{12a} and for
2d quantum gravity
coupled to Potts models \cite{13} yield $d \approx 2.7$ or $2.8$;
and an analytical calculation predicts that $d$ lies between $2$ and $3$
for pure 2d quantum gravity \cite{14}.
Thus we fit the specific heat peak to
\begin{equation}
C_{max} = A' N^{\alpha/\nu d} +B'
\label{e21}
\end{equation}
for our three values of $N$
obtaining $\alpha/\nu d = 0.12(2)$. Thus a first order phase transition
is apparently ruled out. One could argue that our systems are too small to see
the true
scaling of $C_{max}$, however Renken and Kogut managed to see its first order
scaling in their model on systems of size $576$ and $1024$ \cite{7}.
It should be noted, however, that they carried out their simulations
on a regular, fixed lattice rather than the highly disordered lattice we are
using, so the different behavior we observe might be due to
the lattice rather than the absence of the gaussian term.

If we assume that the phase transition is second order and use
the scaling relation
\begin{equation}
\nu d = 2 - \alpha
\label{e22}
\end{equation}
we obtain $\alpha = 0.20(3)$ and $\nu d = 1.8(3)$.
Now, from our measurements of the correlation length, we can obtain $\nu$
by fitting
\begin{equation}
\xi \sim |\lambda_c-\lambda|^{-\nu}
\label{e22}
\end{equation}
for $\lambda < \lambda_c$, with $\lambda_c = 2.35$. This value for
$\lambda_c$ is obtained from the peak in the susceptibility, shown in Fig. 4.
This leads to $\nu = 0.7(1)$.
Therefore we estimate $d = 2.6(8)$. Unfortunately the error here is rather
large but nevertheless the numbers are consistent.
As a final check we could assume $\alpha = 0$ (as is the case for the
phase transition in the usual $2d$ Ising model), then $\nu d = 2$
so we eliminate one source of error and obtain $d = 2.9(4)$.

At this point we could tentatively conclude that our data is consistent with
the phase transition being second order. However, it is
interesting to note similarities
with data from numerical simulations of (gaussian) spin glasses which typically
have a rather broad peak in the specific heat and a susceptibility
peak at lower temperature \cite{15} (which corresponds to
{\it higher} $\lambda$).
If we look at Fig.3 and Fig.4 we see just this behavior. The very
weak growth of the peak in the specific heat that we have measured
could be a finite size effect masking a negative $\alpha$
(in the Sherrington/Kirkpatrick spin glass model $\alpha=-1$ \cite{15})
\footnote{Recent very high statistics
simulations of a gaussian plus extrinsic curvature-squared action
in three dimensions \cite{6} have also found very little, if any, growth in
the peak.}.
Moreover, at the phase transition in our model we have competing interactions
which typically leads to spin-glass type behavior. As these
interactions are still present in the original model with
gaussian plus extrinsic curvature-squared terms this suggests that the
low $\lambda$, crumpled phase in this model could conceivably be spin-glass
like
rather than paramagnetic. A simulation by Heermann \cite{17} has, in fact,
found that the radius of gyration $X2$ is non self-averaging in the
crumpled phase, a behavior found in some observables in spin glasses.

To summarize the numerical results and speculations: we have simulated
an effective model for crumpling in two dimensions in which the $X$ variables
are
replaced by an antiferromagnetic term that mimics their effect on the normals.
We have found what appears to be a weak second (or possibly higher) order
transition, rather than the first order transition of \cite{7}, which retained
the $X$s. We have pointed out that this might be due to the differing
underlying
fixed meshes in the simulations rather than the absence of the $X$ variables.
We have also speculated that the competing nature of the interactions at
the transition may give a spin-glass like phase at low $\lambda$ rather
than a paramagnetic phase and presented some numerical support for this view.

It would be interesting to repeat the simulation on a fixed regular mesh
to compare directly with \cite{7} and to extend the effective model
work to the more realistic case of a surface embedded in three dimensions,
where one would employ continuous spins. The nature of the low
$\lambda$ (``crumpled'') phase also merits further investigation  for
both the effective model and the original action.

This work was supported in part by NATO collaborative research grant CRG910091.
CFB is supported by DOE under contract DE-AC02-86ER40253 and by AFOSR Grant
AFOSR-89-0422.

\vfill
\eject

\vfill
\centerline{\bf Figure Captions}
\begin{description}
\item[Fig. 1.]
Energy for the N=1000 simulation.
\item[Fig. 2.]
Magnetization for all three (N=100,500 and 1000) simulations.
\item[Fig. 3.]
Specific heat for all three simulations.
\item[Fig. 4.]
Susceptibility for all three simulations.
\end{description}
\end{document}